\documentclass[conference,a4paper]{IEEEtran}
\usepackage{amssymb}
\usepackage{xcolor}
\usepackage{amsthm}
\usepackage{mathtools}
\usepackage{siunitx}
\usepackage{cuted}
\usepackage{graphicx}
\usepackage{subcaption}
\usepackage{epstopdf}
\usepackage{tabularx}
\usepackage{multirow}
\usepackage{booktabs}
\usepackage{diagbox}
\usepackage{cuted}
\usepackage{lipsum}
\usepackage[noadjust]{cite}

\usepackage[english]{babel}
\usepackage[utf8]{inputenc}
\usepackage{algorithm}
\usepackage{algpseudocode}
\usepackage{amsmath}

\usepackage{graphicx}
\usepackage{threeparttable} 
\usepackage{caption}

\usepackage{ulem}

\makeatletter
\newcommand{\removelatexerror} {\let\@latex@error\@gobble}
\makeatother

\newif\iftrackrivision

\iftrackrivision

\else

\fi

\renewcommand{\emph}[1]{\textit{#1}}

\IEEEoverridecommandlockouts

\begin{document}

\title{Massive Twinning to Enhance Emergent Intelligence}

\author{
	\IEEEauthorblockN{Siyu Yuan\IEEEauthorrefmark{1},~Bin~Han\IEEEauthorrefmark{1},~Dennis Krummacker\IEEEauthorrefmark{2},~and~Hans~D.~Schotten\IEEEauthorrefmark{1}\IEEEauthorrefmark{2}}
	\IEEEauthorblockA{
	\IEEEauthorrefmark{1}Division of Wireless Communications and Radio Positioning (WICON), Technische Universit\"at Kaiserslautern\\
	\IEEEauthorrefmark{2}Research Department Intelligent Networks, German Research Center of Artificial Intelligence (DFKI)\\
	\IEEEauthorrefmark{1}\{yuan $\vert$ bin.han $\vert$ schotten\}@eit.uni-kl.de; \IEEEauthorrefmark{2}\{Dennis.Krummacker $\vert$ Hans\_Dieter.Schotten@dfki.de\}@dfki.de
	}
}


\maketitle

\begin{abstract}
As a complement to conventional AI solutions, emergent intelligence (EI) exhibits competitiveness in 6G IIoT scenario for its various outstanding features including robustness, protection to privacy, and scalability. However, despite the low computational complexity, EI is challenged by its high demand of data traffic in massive deployment. We propose to leverage massive twinning, which 6G is envisaged to support, to reduce the data traffic in EI and therewith enhance its performance.
\end{abstract}


\IEEEpeerreviewmaketitle

\section{Introduction}\label{sec:intro}
As backbone of Industry 4.0 (I4.0), industrial wireless network (IWN) and Industrial Internet-of-Things (IIoT) have been attracting significant research interests over the past years~\cite{tange2020systematic}. This technological trend is reflected by the introduction of massive machine-type communication (mMTC) and ultra-reliable low-latency communication (URLLC) use cases in the 5G wireless networks. Pushing the trend forth, efforts to define the 6G generally consider it an essential and characterizing feature to deliver pervasive intelligent services in industrial scenarios~\cite{dang2020should}. Especially, ubiquitous deployments of artificial intelligence (AI) and digital twins (DTs) are envisaged~\cite{uusitalo20216g}.

It is widely suggested to create and maintain a DT for almost every physical entity (including humans) in the involved industrial process/scenario, and to exploit this massive twinning paradigm in AI applications by means of provisioning, aggregating, and analyzing data~\cite{rathore2021role}. However, this approach introduces notable security risks. Relying on centralized decision maker and globally shared model, conventional AI solutions are fragile against cyber-attacks at the decision making unit, the model, or the database that supports model training. 
Therefore, the deep integration of AI in I4.0 is widely recognized to form a new attack surface~\cite{becue2021artificial}. To resolve privacy leakage in AI systems~\cite{li2020federated}, distributed AI solutions such as Federated Learning (FL) have been emerging. By distributing the AI engine from the central controller to every user, it lets no raw user data but only the model parameters be exchanged between users and server, and therewith reduces data exposing. Nevertheless, due to its model-based nature, FL remains sensitive to malicious model manipulation, which can be realized by attacking techniques such like fake data injection. Furthermore, even FL may still fail to exclude the risk of privacy leakage, e.g. when challenged by information stealing attacks that leverage generative adversarial networks~\cite{sun2021decentralized}. 

As an alternative solution, emergent intelligence (EI) exhibits some outstanding features. Especially, it relies on no explicit system-level model at any local agent, but only the spontaneous emergence of complex behavior patterns from the interaction among massive agents with simple logic. This model-less nature and the ad hoc topology grant EI with excellent privacy friendliness and appreciable robustness against local failures. Furthermore, EI is known to have a good scalability, with its effectiveness benefiting from the number of agents~\cite{olorunda2008measuring}. However, the performance of EI highly relies on the information exchange among  agents. Thus, the deployment of EI in practical IWN scenario is challenged by lossy channels, significant signaling overhead, and limited radio resources. 

To address these issues, in this paper we propose to leverage in EI applications the massive twinning approach. It migrates both the real-time data and the decision making logic from mobile devices to the multi-access edge computing (MEC) server, and therewith significantly reduces the traffic load while enhancing the reliability. To demonstrate this idea, we consider an unmanned aerial vehicles (UAV) use case, where multiple UAVs jointly localize and assemble at a target location, while each UAV can only measure its own position and distance to the target. We implement a partical swarm optimization (PSO) algorithm to solve this task as a typical instance of EI approaches, and tested its performance in different communication schemes under the same simplified resource constraints, through which we are able to demonstrate the gain EI can obtain from the massive twinning paradigm.

The remainder of this paper is organized as follows: We setup the problem in Sec.~\ref{sec:setup}, for which we propose a swarm solution and specify it to different radio scenarios in Sec.~\ref{sec:approaches}. The performance is evaluated by simulations, which we present and discuss in Sec.~\ref{sec:simulations}. To the end, we make some additional discussion about technical details in Sec.~\ref{sec:discussions}, before closing this paper with our conclusion and outlooks in Sec.~\ref{sec:conclusion}.

\section{System Model}\label{sec:setup}
\begin{figure}[!htbp]
	\centering
	\includegraphics[width=.8\linewidth]{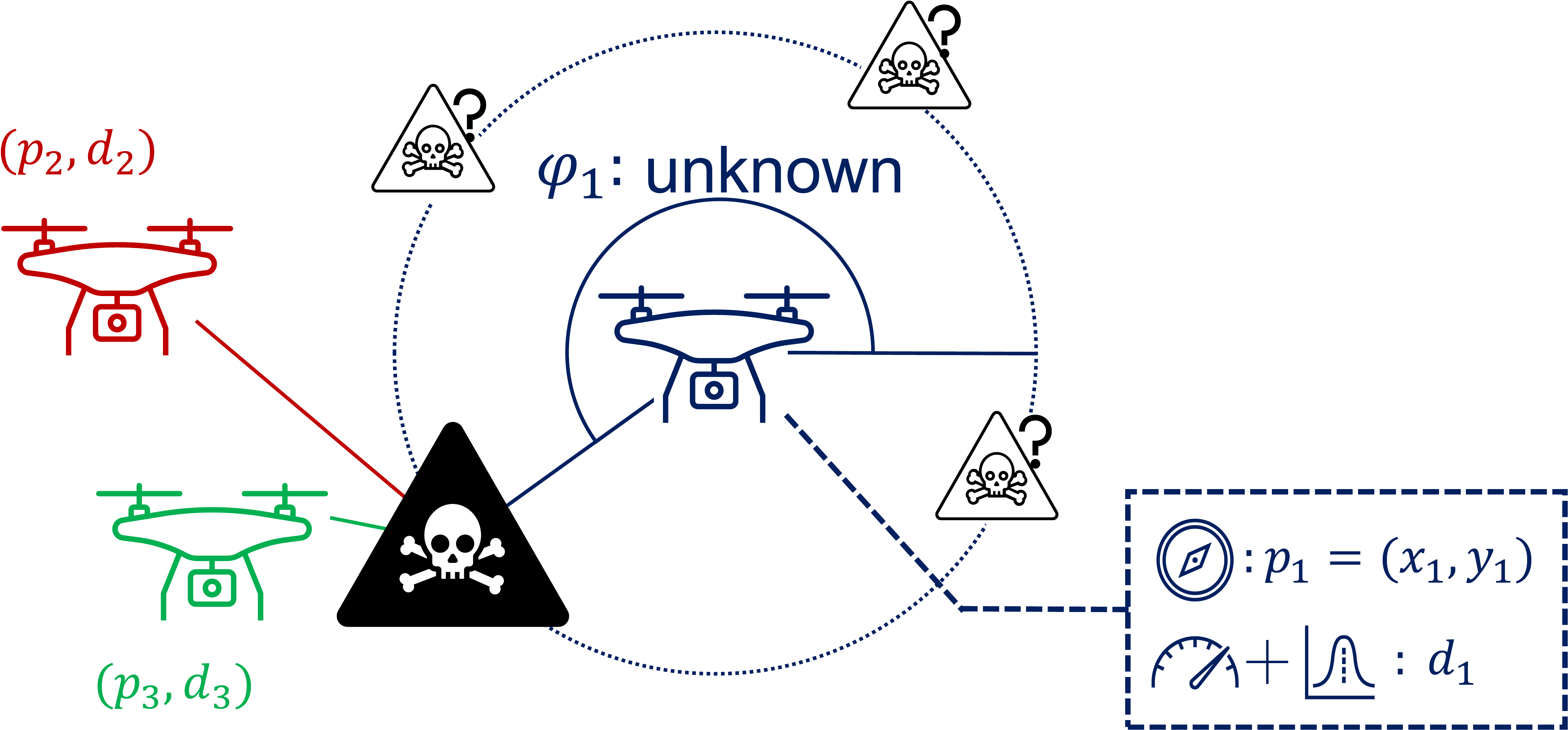}
	\caption{Multi-UAV-based handling of chemical spills}
	\label{fig:application}
\end{figure}

We consider an application of emergency reaction to chemical spills, where multiple UAVs are deployed at different locations over a chemical manufacturing site, as illustrated in Fig.~\ref{fig:application}. Supposed to localize the spill spot and arrive there to take all necessary repairing measures, each UAV (also referred to as an \textit{agent} hereafter) is equipped with not only a positioning module to obtain its own position, but also sensors to measure the chemical densities in the air, from which it can estimate its distance to the spill spot according to the diffusion model. However, due to the isotropic diffusion of chemicals in the air, no single agent is capable of estimating the relative direction it is distanced from the spill spot. Thus, a joint localization and routing must be carried out, which exploits the measured data of multiple agents.

To simplify the demonstration and focus on evaluating the impact of massive twinning in EI applications, in this work we consider no obstacle that blocks the sensing, communication, or movement of the agents. The agents are initially distributed at random locations, and act in an iterative manner. In every round, each agent first measures its own position and distance to the spill, then communicate with the other agents to exchange this information, and moves to the next position regarding the collected information. When every agent has reached its new position, the round is closed and a new round starts. In this way, the agents gradually converge to the spill.

Concerning the practical physical constraints and different ways of communication, we make the following assumptions: \textit{i}) The speed $\vert v_i^t\vert$ of each agent $i$ in the agent set $\mathcal{I}$ in every round $t$ is restricted to a constant limit $\vert v_\text{max}\vert$; \textit{ii}) The distance $d_i^t=\vert p_i^t-p_\text{des}\vert$ between agent $i$ and spill in round $t$ cannot be precisely measured, but only inaccurately estimated as $\hat{d}_i^t=d_i^t+e_i^t$ where $e_i^t\sim\mathcal{N}(0,\sigma^2)$  for all $t\in\mathbb{N}^+$; and \textit{iii}) Every agent $i$ is not guaranteed able to communicate with all other agents, but only with a subset $\mathcal{I}_i^t\subseteq\mathcal{I}$ of agents. 

\section{Approaches}\label{sec:approaches}
\subsection{SWARM}
As the most significant category of EI solutions, swarm intelligence (SI) provides a variety of classical algorithms to solve tasks of such kind. In this work, we invoke the the most classical PSO \cite{kennedy1995particle} as our swarm convergence method. In every round $t\in\mathbb{N^+}$, every agent $i\in\mathcal{I}$ updates in each round the historical best positions of its own and of the entire swarm, and therewith individually makes its own decision of next move : 
\begin{align}
v^{t+1}_{i}&=v^{t}_{i} + c_{1}r_{1}\left(p^{t}_{i,\text{best}} - p^{t}_{i}\right)+ c_{2}r_{2}\left(p^{t}_\text{best} - p^{t}_{i}\right), \label{eq:velocity}\\
p^{t+1}_{i}&=p^{t}_{i} + v^{t+1}_{i},\label{eq:loc_update}
\end{align}
where $p^{t}_{i,\text{best}}$ is the best historical till round $t$ position of agent $i$ that minimizes $d_i$, and $p^{t}_\text{best}$ the best historical position of the entire swarm. The constant coefficients $c_1$ and $c_2$ specify the weights of local knowledge and swarm knowledge, respectively. The pair of random numbers $(r_1,r_2)\sim\mathcal{N}^2$ introduce randomness into the update to enhance the convergence performance. Recalling our assumption of limited communication capability, in our studied case, $p^t_\text{best}$ that can only be obtained from the aggregated knowledge of all agents in $\mathcal{I}$ is not guaranteed available for ever individual $i$. Thus, we replace it with a ``partial'' knowledge $g_{i,\text{best}}$, which is the best historical position of all agents in $\mathcal{I}_i^t$. Taking the practical conditions of speed limit and inaccurate distance also into account, we adopt the PSO algorithm on our application as described in Algorithm~\ref{alg:sketch}.

\begin{algorithm}[!htbp]
		\caption{The modified PSO algorithm of our proposal}
		\label{alg:sketch}
		\begin{algorithmic}[1]
			\scriptsize
			\State \textbf{Input} $\mathcal{I}, \{p_i^1,~\forall i\in\mathcal{I}\}, p_\text{des}, \vert v_\text{max} \vert, c_1, c_2, T$ 
			\For {$i=1:I$}
				\State{Update $\hat{d}_i^1$}
				\State{$d_{i,\text{best}}=\hat{d}_i^1$, $p_{i,\text{best}}=p_i^1$, $d_{\mathcal{I}_i,\text{best}}=\hat{d}_i^1$, $p_{\mathcal{I}_i,\text{best}}=p_i^1$}
			\EndFor
			\For{$t = 1: T$}
			\For {$i=1:I$}
			\State{Update $\mathcal{I}_i^t$}\label{line:update_neighbors}
			\For {$j \in \mathcal{I}_i^t$}
				\State {$i$ inquiries $p_j^t$ and $\hat{d}_j^t$ from $j$}\label{line:communication}
				\If{$\hat{d}_j^t<g_{i,\text{best}}$}
				\State{$g_{i,\text{best}}=\hat{d}_j^t$}
				\EndIf
			\EndFor
			\If{$\hat{d}_i^t<d_{i,\text{best}}$}
				\State{$d_{i,\text{best}}=\hat{d}_i^t$, $p_{i,\text{best}}=p_i^t$}
			\EndIf
			\State{Generate random numbers $r_1,r_2$}
			\State {$v^{t+1}_{i}=v^{t}_{i} + c_{1}r_{1}\left(p^{t}_{i,\text{best}} - p^{t}_{i}\right)+ c_{2}r_{2}\left(p^{t}_{\mathcal{I}_i,\text{best}} - p^{t}_{i}\right)$}
			\If {$\vert v_i^{t+1}\vert>\vert v_\text{max}\vert$}
			\State {$v_i^{t+1} = v_i^{t+1}\times \frac{\vert v_{\text{max}}\vert}{\vert v_i^{t+1}\vert }$}
			\EndIf
			\State {$p_i^{t+1} = p_i^t+v_i^{t+1}$}
			\State {Generate $e_i^{t+1}$}
			\State {$d_i^{t+1} = p_i^{t+1}-p_\text{dest}+e_i^{t+1}$}
			\EndFor
			\EndFor
		\end{algorithmic}
	\end{algorithm}

\subsection{Networking}
Algorithm~\ref{alg:sketch} generically requires to setup communication links and transmit data between different agents, which in practice highly depends on the networking design. Hereafter, we propose three candidate schemes.

\subsubsection{D2D}
\begin{figure}[!htpb]
	\centering
	\includegraphics[width=.8\linewidth]{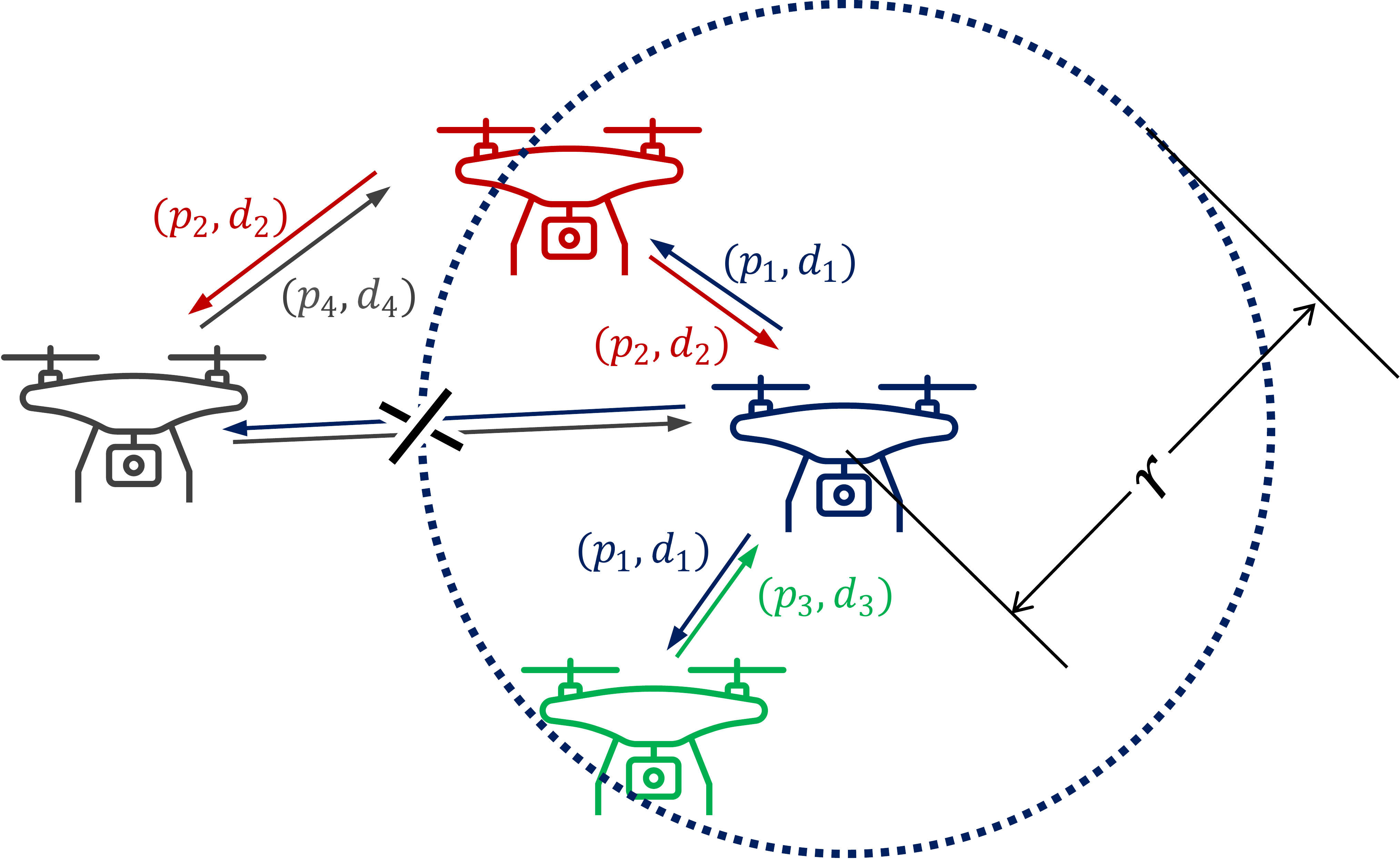}
	\caption{The D2D-based UAV networking topology}
	\label{fig:d2d_networking}
\end{figure}

Device-to-device (D2D) communication is known to be radio resource efficient by means of resolving access collisions, reducing end-to-end latency, and raising link reliability~\cite{HSH+2019d2d,JWHS2018applying}. However, restricted by the energy and interference management, D2D technologies are usually limited in coverage, as illustrated shown in Fig.~\ref{fig:d2d_networking}. With a sufficient coverage radius $r$, it takes $2I(I-1)$ successful radio transmissions to fully exchange information among $I$ agents.

\subsubsection{Classical cellular solution}
\begin{figure}[!htpb]
	\centering
	\includegraphics[width=.7\linewidth]{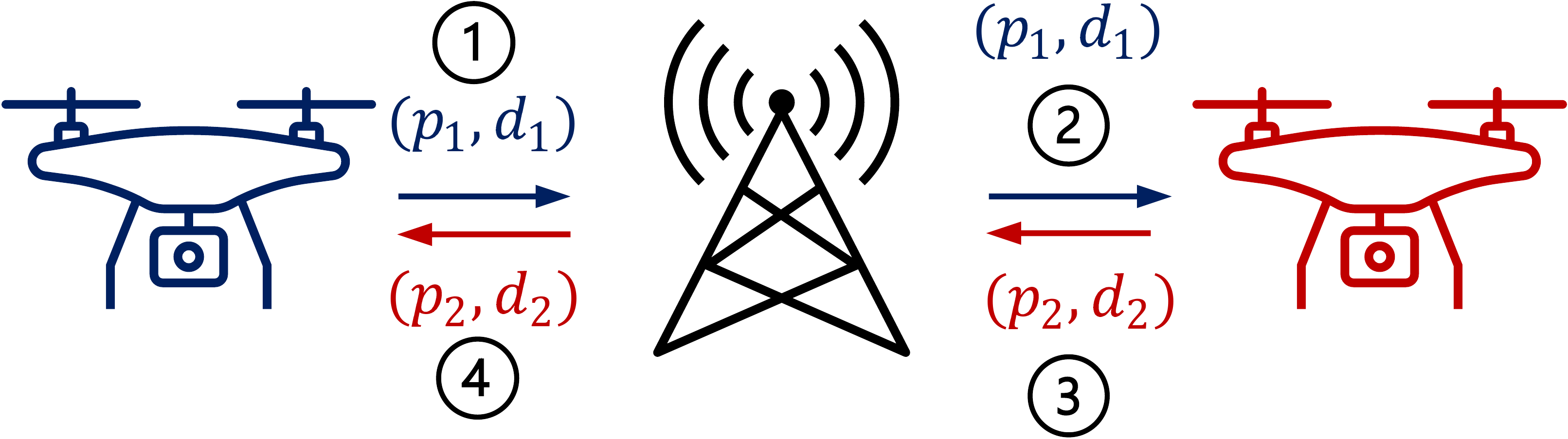}
	\caption{The classical cellular UAV networking topology}
	\label{fig:bs_networking}
\end{figure}

An alternative solution is to rely on a cellular base station to forward messages between agents, as illustrated in Fig.~\ref{fig:bs_networking}: Agent 1 (blue) initiates the session by sending to BS in uplink its information and the identification of an target agent (agent 2 in red). Upon a successful reception, the BS forwards this information to agent 2 in downlink. If agent 2 successfully decodes this message, it responses by sending its information back to agent 1. Thus, it takes $4I(I-1)$ successful radio transmissions to fully exchange information among $I$ agents. 

\subsubsection{Digital twin enhancement}
\begin{figure}[!htpb]
	\centering
	\includegraphics[width=.9\linewidth]{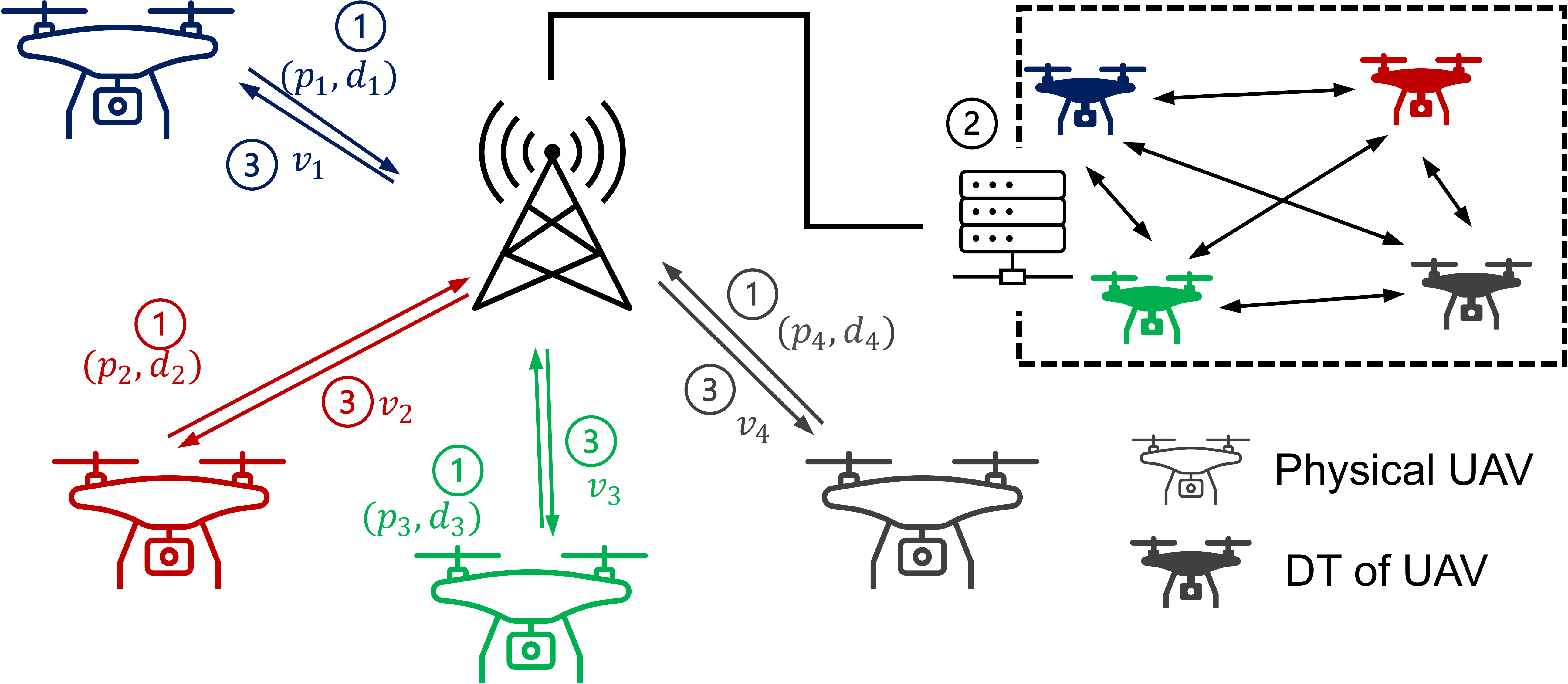}
	\caption{Cellular UAV networking with massive twinning}
	\label{fig:dt_networking}
\end{figure}

The radio traffic can be reduced by deploying massive twinning for a better integration in IIoT scenarios. As illustrated in Fig.~\ref{fig:dt_networking}, each UAV has a DT at the MEC server. In each round, position and distance information are offloaded by every UAV to its DT (step 1), and shared among the DTs without radio trafficr (step 2). Every DT then makes the decision for its next move and send it back to the UAV (step 3). Thus, it takes $2I$ successful radio transmissions to fully exchange information among $I$ agents.

\section{Simulations}\label{sec:simulations}
\begin{figure}[!htpb]
    \centering
    \includegraphics[width=.5\linewidth]{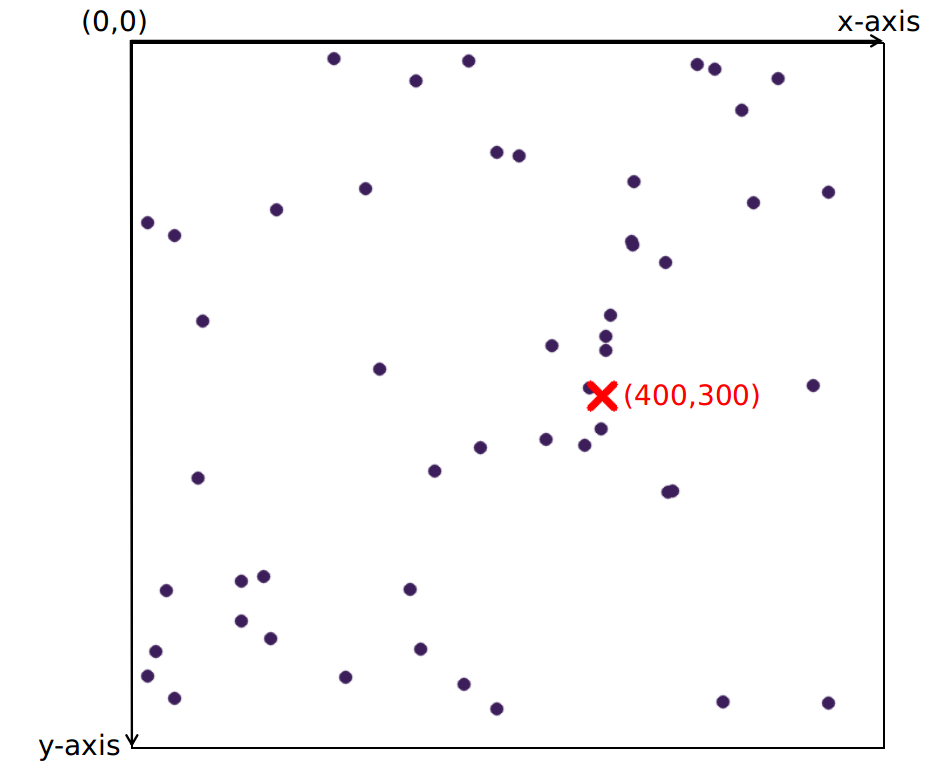}
    \caption{The map of simulation scenario}
    \label{fig:SimGround}
\end{figure}

To evaluate the performance of EI upon various networking scheme, we carried out numerical simulations regarding the scenario in Fig.~\ref{fig:SimGround}. UAVs are uniformly randomly over a \SI{640}{}$\times$\SI{600}{\meter^2} map, and the spill is at $p_{\text{des}}=(\SI{400}{\meter},\SI{300}{\meter})$. All agents' maximal speed are limited to $\vert v_\text{max} \vert=\SI{5}{\meter/}\text{round}$, and the deviation of distance sensing error $\sigma = \SI{1}{\meter}$.

\subsection{D2D vs. BS}
\begin{figure}[!htpb]
	\centering
	\begin{subfigure}[t]{\linewidth}
		\centering
		 \includegraphics[width=.7\linewidth]{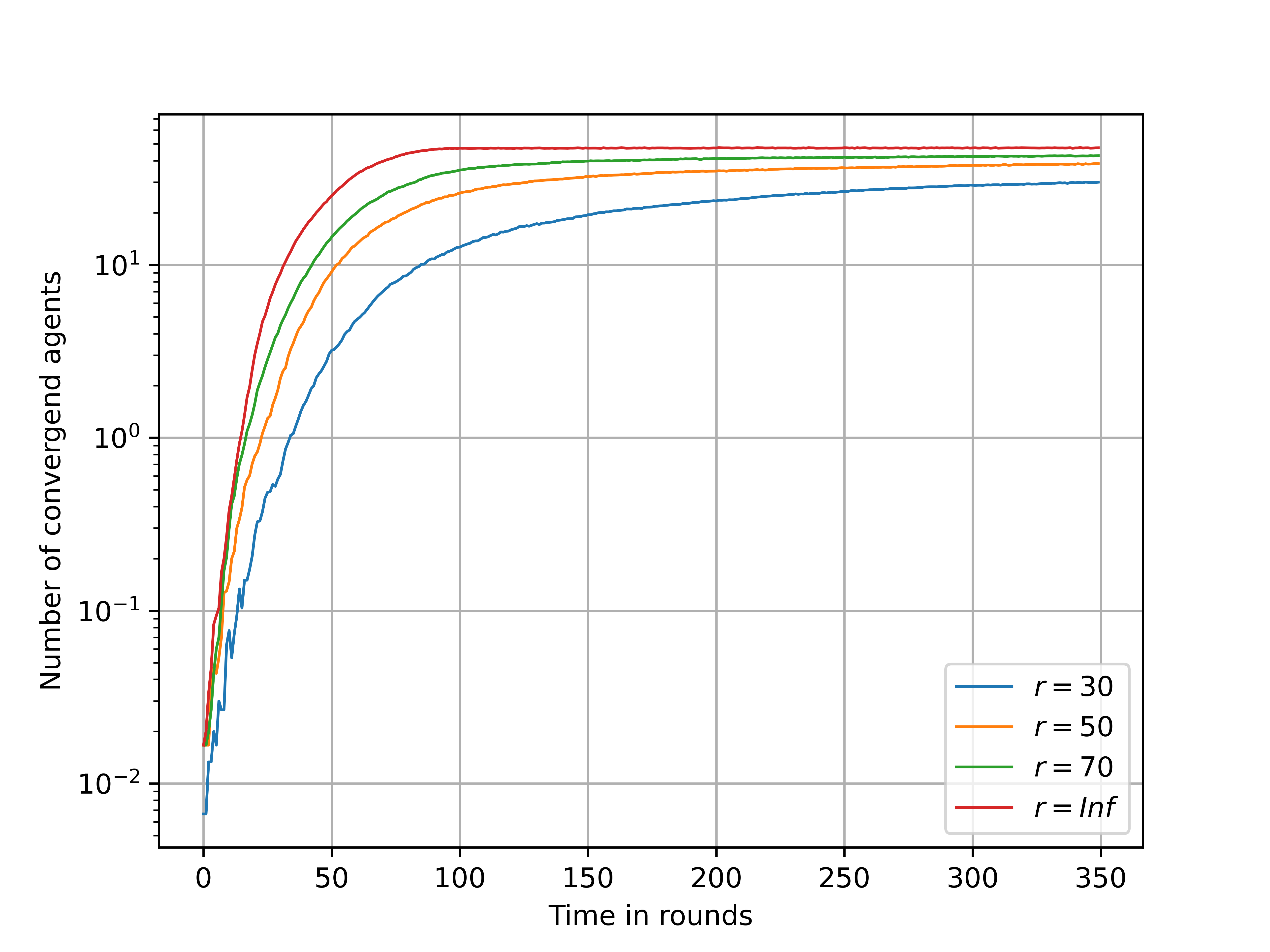}
		 \caption{D2D solution}
	\end{subfigure}
	\begin{subfigure}[t]{\linewidth}
		\centering
		\includegraphics[width=.7\linewidth]{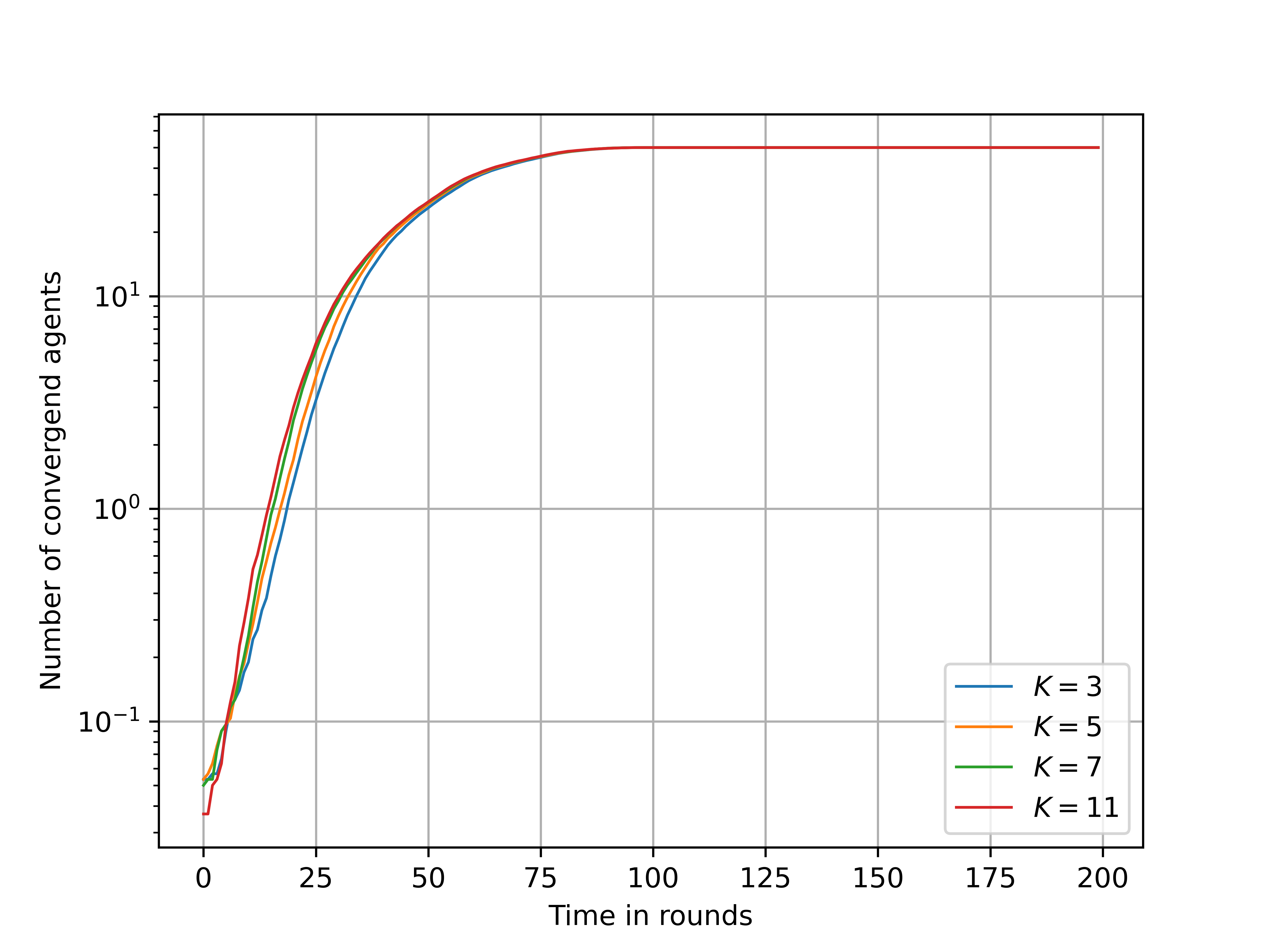}
		 \caption{Classical cellular solution}
	\end{subfigure}
	\caption{Convergence of Alg.~\ref{alg:sketch} upon networking schemes.}
	\label{fig:d2d_vs_bs}
\end{figure}

First we investigated the convergence performance in setups of D2D-based and classical cellular networking schemes. For the D2D setup, we assumed every agent to be capable of successfully communicating \emph{with and only with} all neighbor agents within the coverage radius $r$. For the D2D setup, we assumed that in every individual round, every agent is able to initiate and complete communication with $K$ other randomly selected agents, without any error or failure.  In both cases, $50$ agents were deployed. We carried out Monte-Carlo tests with different values of $r$ and $K$, respectively, repeating under every specification 300 runs, each time with i.i.d. initial agent locations. The  results are depicted in Fig.~\ref{fig:d2d_vs_bs}, showing a significant degradation of convergence in D2D scheme with lower coverage, and a guaranteed converge under BS assistance even with very sparse connection among the agents. Despite of the highly simplified communication models applied here, the results sufficiently convince us to claim that the BS-based solution generally outperforms the D2D-based one in convergence. The reason is that the D2D-based networking allows every agent only to obtain a ``regional'' knowledge from neighbor agents that are close to itself, which delivers only marginal extra information in addition to the agent's own knowledge. In contrast, the BS-based solution allows every agent to obtain a partial but yet global knowledge from agents all over the map, which contains more mutual information due to the spatial diversity among agents.

%


\subsection{DT enhancement for scalability}
\begin{figure}[!htpb]
	\centering
	\begin{subfigure}[t]{\linewidth}
		\centering
		\includegraphics[width=.6\linewidth]{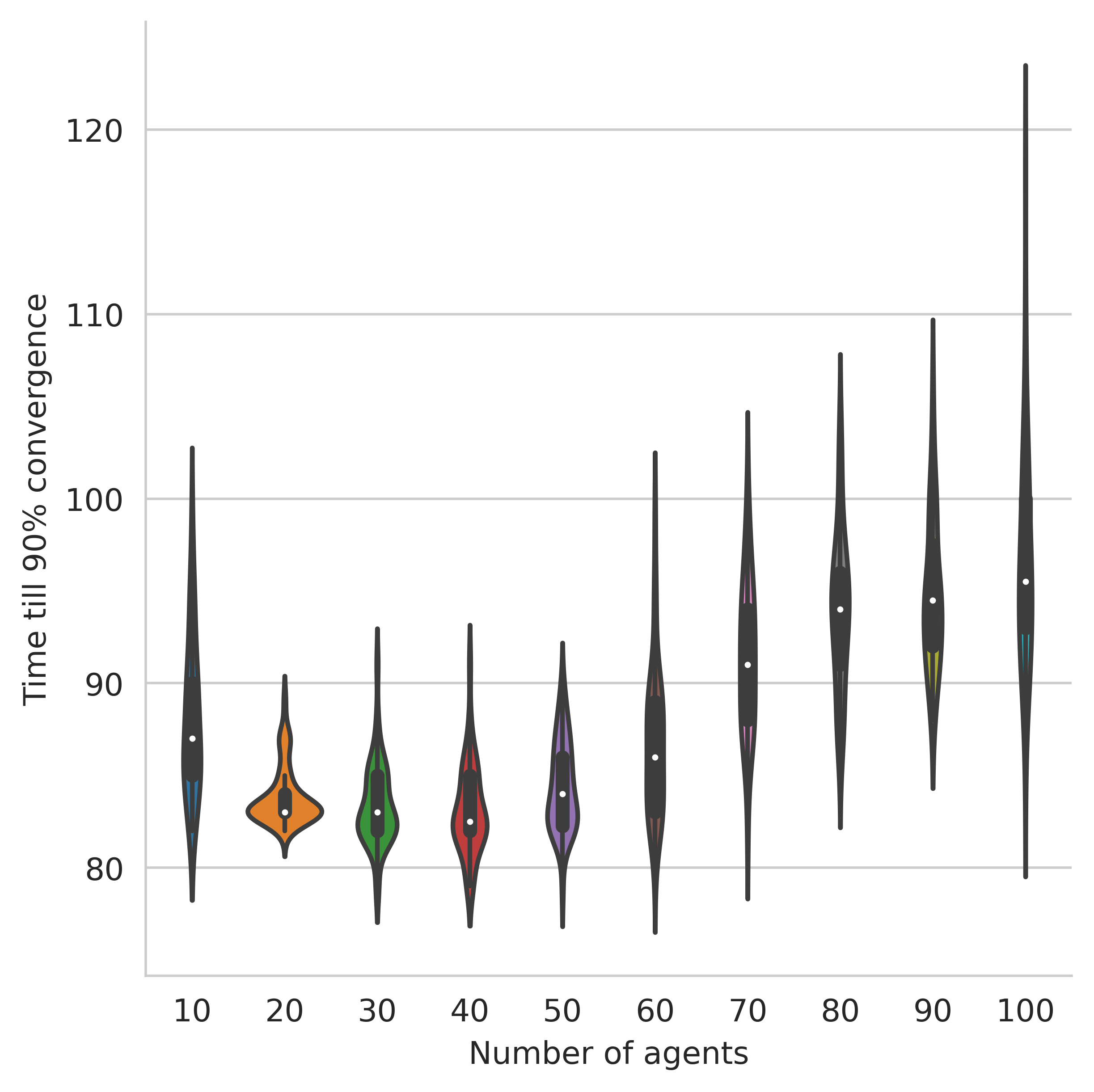}
		\caption{Classical cellular solution}
	\end{subfigure}
	\begin{subfigure}[t]{\linewidth}
		\centering
		\includegraphics[width=.6\linewidth]{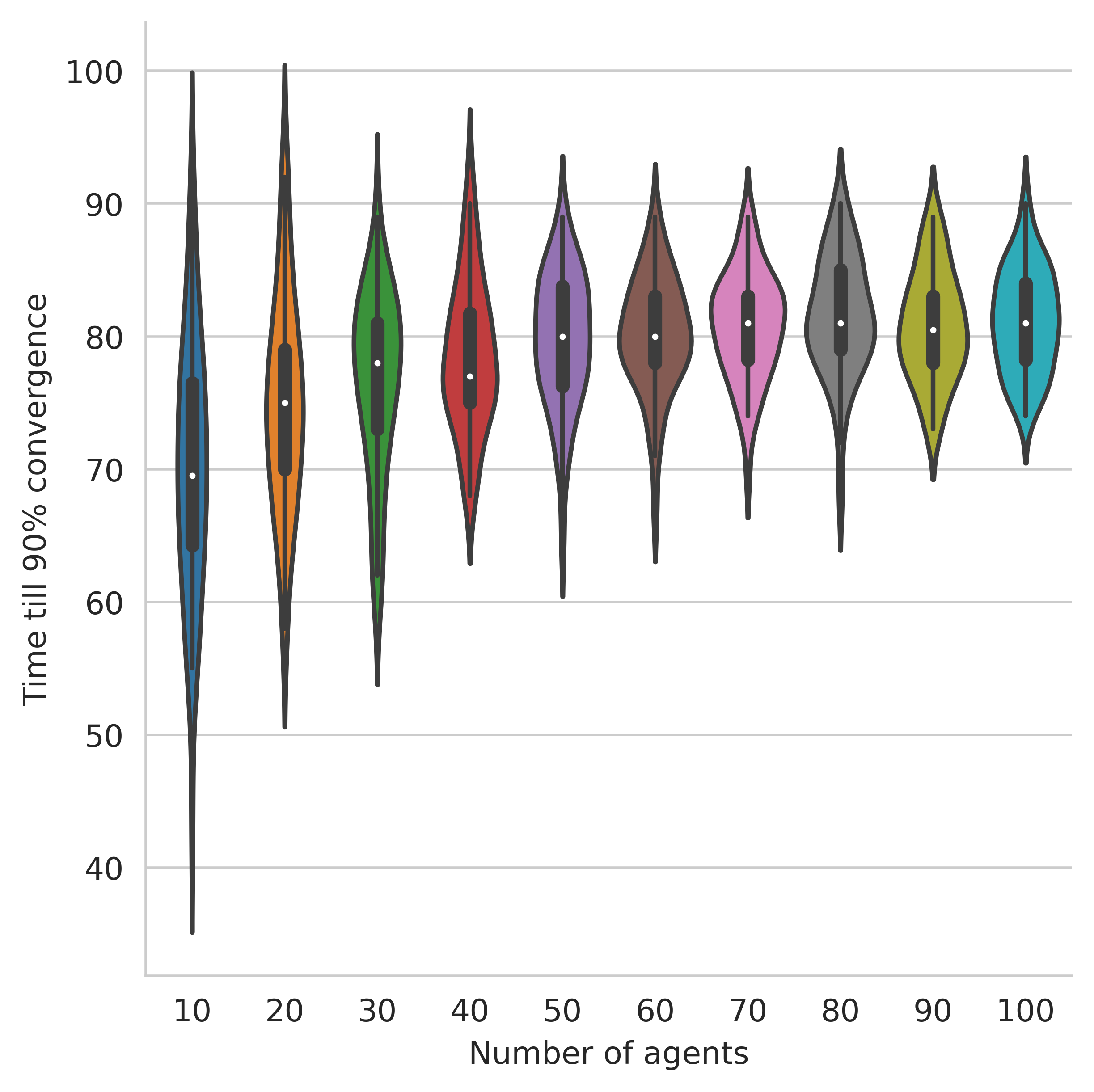}
		\caption{Enhanced with massive twinning}
	\end{subfigure}
	\caption{Convergence of Alg.~\ref{alg:sketch} upon agent number. }
	\label{fig:comparison_bs_dt}
\end{figure}

The time complexity of communication in classical cellular scheme ($\mathcal{O}(I^2)$) and the DT-based solution ($\mathcal{O}(I)$) is implying a difference in their system scalability. To evaluate this impact, we compared their convergence performance with various number of agents but under a shared constraint of up to $1000$ radio transmissions for the entire system. The test was repeated 50 times for each specification with i.i.d. random initial agent locations. As shown in Fig.~\ref{fig:comparison_bs_dt}, the massive twinning enhancement brings a significant gain in converging speed under \textit{all} specifications. 

\section{Discussion}\label{sec:discussions}
As a pioneering demonstrative work, this study considers highly simplified communication models, where neither bit errors nor access collisions are taken into account. In realistic radio scenarios, the communications suffer from random latency and packet losses, which reduce the quality of communication and therefore degrade the convergence performance of the swarm algorithm. Nevertheless, since our proposed DT-based solution relies less on the radio channels than the classical cellular approach, it will certainly suffer less from these effects than the latter, and the performance gain provided by DT shall be therefore larger than observed in this study.

\section{Conclusion and Outlooks}\label{sec:conclusion}
In this paper, we have investigated the swarm algorithm as an instance of EI in the context of future IIoT, studying how it can benefit from the massive twinning paradigm that will be supported by 6G. Based on some simplified scenarios, we have demonstrated that the deployment of DT will enhance the performance and scalability of EI solutions by effectively reducing the demand of radio traffic. Looking forward to future research, we consider it interesting and essential to evaluate our proposed solution in more realistic radio scenarios and specified to practical communication systems.

\section*{Acknowledgement}
This work is partly funded by the European Commission through the H2020 project Hexa-X (GA no. 101015956).

\ifCLASSOPTIONcaptionsoff
  \newpage
\fi


\bibliographystyle{IEEEtran}
\bibliography{CiteTheses}

	



\end{document}